# A Similarity Analysis for Heat Transfer in Newtonian and Power Law Fluids Using the Instantaneous Wall Shear Stress

Trinh, Khanh Tuoc

*K.T.Trinh@massey.ac.nz*
B.H.P. Wilkinson

*B.Wilkinson@massey.ac.nz*
Institute of Food Nutrition and Human Health

Massey University, New Zealand
N.K.Kiaka
*nkiaka@appsci.unitech.ac.pg*
Department of Applied sciences
The Papua New Guinea University of technology

## Abstract

This paper presents a technique that collapses existing experimental data of heat transfer in pipe flow of Newtonian and power law fluids into a single master curve. It also discusses the theoretical basis of heat, mass and momentum analogies and the implications of the present analysis to visualisations of turbulence.

## 1   Introduction

The study of heat and mass transfer has been dominated from an early stage by the similar form of the equations of heat, mass and momentum. Following Boussinesq (1877), the transport flux (e.g. of heat) can be defined in terms of an eddy viscosity

$$q = -(k + \rho E_h)\frac{d\theta}{dy} \tag{1}$$

where   $\theta$   is the temperature

   $y$   the normal distance from the wall

   $k$   the thermal conductivity

   $E_h$   the eddy thermal diffusivity

   $q$   the rate of heat transfer flux

   $\rho$   the fluid density

Equation (1) may be rearranged as

$$\theta^+ = \int_0^{y^+} \frac{q/q_w}{1 + E_h/k} dy^+ \tag{2}$$

which is very similar to the equation for momentum transport

$$U^+ = \int_0^{y^+} \frac{\tau/\tau_w}{1 + E_v/\nu} dy^+ \tag{3}$$

Where $U^+ = U/u_*$, $y^+ = yu_*\rho/\mu = yu_*/\nu$ and $\theta^+ = \rho C_p (\theta - \theta_w) u_*/q_w$ have been normalised with the friction velocity $u_* = \sqrt{\tau_w/\rho}$ and the fluid apparent viscosity $\mu$. The suffices w refer to the parameters at the wall, $\nu$ momentum, h heat and $\tau$, $E_v$ are the shear stress and eddy diffusivity for momentum respectively.

## 1.1 Reynolds' analogy

Reynolds (1874) was the first to propose a formal analogy between heat, mass and momentum transfer expressed

$$St = \frac{f}{2} \tag{4}$$

where

   $St = \dfrac{h}{\rho C_p V}$   is called the Stanton number,

   $f = \dfrac{2\tau_w}{\rho V^2}$   the friction factor,

$C_p$ is the thermal capacity and

$V$ the average fluid velocity (5)

This requires that the normalised velocity and temperature profiles be the same (Bird et al., 1960) p382.

$$\frac{d\theta^+}{dy^+} = \frac{dU^+}{dy^+} \qquad (6)$$

It is normally assumed that Reynolds analogy implies two conditions

$$q/q_w = \tau/\tau_w \qquad (7)$$

$$\Pr_t = E_h/E_v = 1 \qquad (8)$$

where $\Pr_t$ is called the turbulent Prandtl number. Equations (77) and (88) are at odds with experimental evidence. These are the paradoxes of the Reynolds analogy. The distributions of heat flux and shear stress in turbulent pipe flow are shown in Figure 11 clearly are not equal. The distribution of shear stress is linear and unique for all Reynolds numbers but the distribution of heat fluxes is dependent on both the Reynolds and Prandtl number.

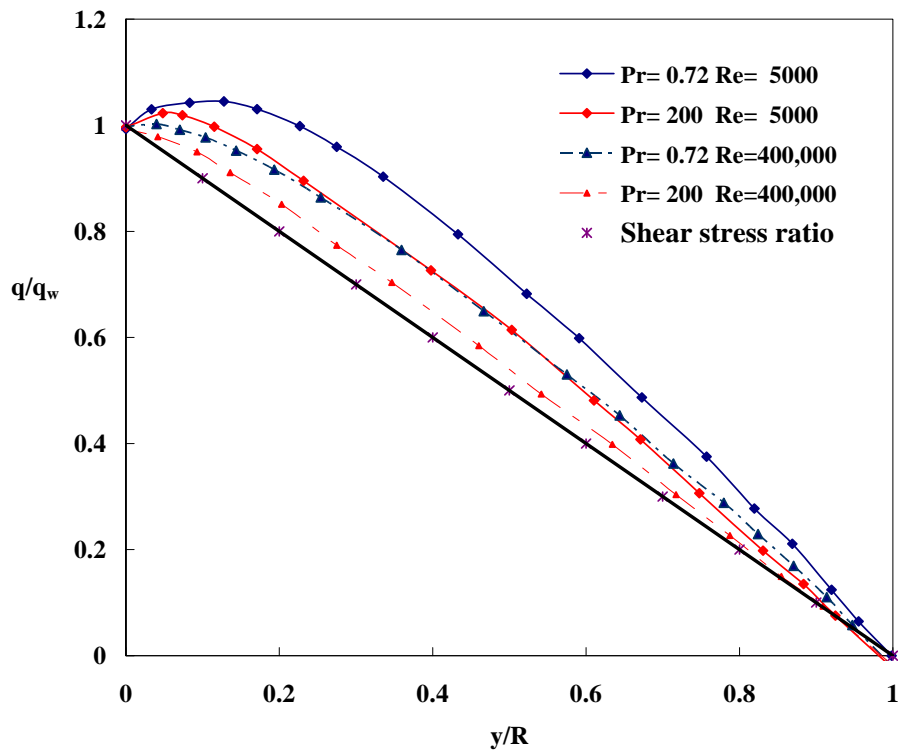

Figure 1. Distribution of shear stress and heat flux in turbulent pipe flow. From Hinze (1959).

Similarly, many workers have shown that the turbulent Prandtl number $Pr_t$ is not unity (Blom and deVries, 1968, McEligot et al., 1976, Malhotra and Kang, 1984, Kays, 1994, McEligot and Taylor, 1996, Churchill, 2002, Weigand et al., 1997).

The paradoxes of Reynolds' analogy can easily be explained (Trinh, 1969). Equation (66) can be obtained from equations (1) and (22) without recourse to the assumptions in equations (77) and (88) when we apply a number of simplifications that require

$$\frac{E_h}{E_v} = \frac{q/q_w}{\tau/\tau_w}, \quad \frac{E_h}{k} \gg 1 \quad and \quad \frac{E_v}{\nu} \gg 1 \tag{9}$$

Equation (96) applies only when the diffusive contributions to transport are negligible i.e. when both the velocity and temperature profiles follow a log-law as shown in Figure 2. Thus it was realised very early that Reynolds' analogy only applied to the turbulent core and Prandtl (1910) improved it by assuming that it only applied up to the laminar sub-layer.

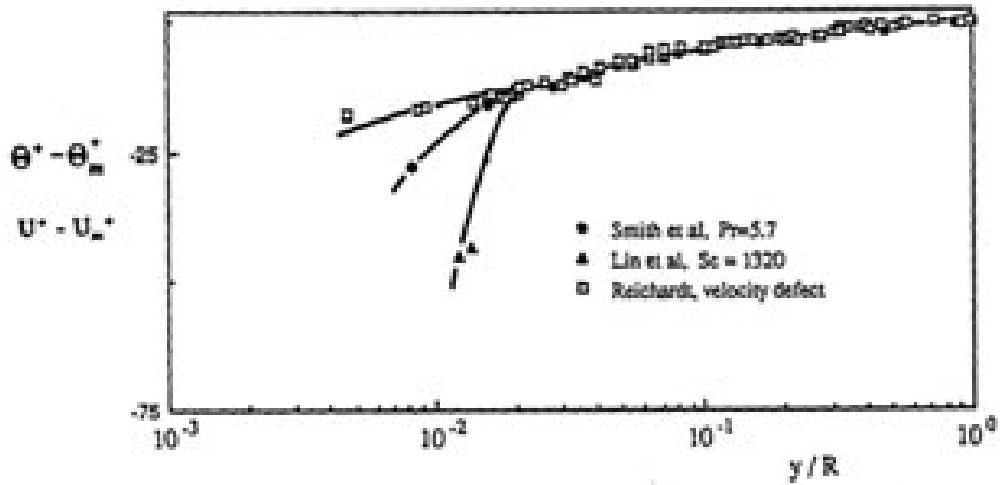

Figure 2  Reynolds' analogy and its range of applicability. Data of (Smith et al., 1967, Lin et al., 1953, Reichardt, 1943).

## 1.2 Analogies between heat, mass and momentum transfer in Newtonian fluids

Since these early days, many other analogies have tried to improve the agreement with experimental data. Three main approaches have been adopted:

1. Empirical correlations, the best known being Colburn's analogy (1933)

$$St = \frac{f}{2} \Pr^{-2/3} \qquad (10)$$

Colburn originally formulated it for pipe flow but subsequent experimental verifications show a discrepancy of about 15% as summarised for example in Bird et al. (1960, p. 400). The agreement for external boundary layer flow is better.

2. Boundary layer theories e.g. (Karman, 1939, Deissler, 1955, Martinelli, 1947, Metzner and Friend, 1958a, Reichardt, 1961, Levich, 1962, Metzner and Friend, 1958b) are based on solutions of equation (21). The list is illustrative and by no means exhaustive of the literature available.

As in momentum transfer, the closure of the models is based on mathematical or physical postulates about the eddy diffusivity $E_h$. This information is often supplied through some experimental measurement of the turbulent Prandtl number. Because of this, boundary layer theories are still referred to as analogies, even though their form has been obtained in a more theoretical framework than the Colburn analogy. In particular Spalding (1961) successfully expanded the velocity near the wall in a Taylor series and showed that the eddy diffusivity scaled with $y^3$. By analogy the eddy thermal diffusivity is assumed to be $E_h \sim y^3$ in many heat transfer studies (Churchill, 1996, 1997).

3. Penetration theories originated with Higbie (1935) who used the equation for unsteady conduction to model the transport process in jets and packed columns.

$$\frac{\partial \theta}{\partial t} = \alpha \frac{\partial^2 \theta}{\partial y^2} \qquad (11)$$

where $\alpha$ is the thermal diffusivity. The well-known solution is

$$q_w = \frac{1}{\sqrt{\pi}} \frac{\Delta \theta_m}{\sqrt{\alpha t_h}} \tag{12}$$

Higbie closed the derivation by assuming that the typical time scale over which equation (12) applies is the contact time

$$t = \frac{x}{U_\infty} \tag{13}$$

where x is the swept length. In an effort to apply Higbie's approach to turbulent transport, Danckwerts (1951) assumed that the surface near the wall is periodically swept clean by eddies penetrating from the bulk stream. The rate of renewal of the surface fluid near the wall is a function of the probability of occurrence of eddies of various frequencies. Danckwerts assumed this probability distribution to be uniform. Subsequent postulates of the surface renewal distributions have been reviewed by Mathpati & Joshi (2007), Pletcher (1988), Ruckenstein (1987), Sideman & Pinczewski (1975). Many of these postulates do not link the assumed distribution of eddies to the improved understanding of the coherent structures or the wall structure but more recent work does e.g. (Fortuin et al., 1992).

Ruckenstein (1968) first attempted to derive a physical model for the distribution function by modelling the eddy as a roll cell which circulates the fluid from the wall to the outer region. The motion close to the wall surface is assumed to obey the laminar transport equation

$$U \frac{\partial \theta}{\partial x} + V \frac{\partial \theta}{\partial y} = \alpha \frac{\partial^2 \theta}{\partial y^2} \tag{14}$$

Ruckenstein calls this state "pseudo-laminar flow" but does not elaborate about the relation between this state and the bursting phenomenon at the wall. Thomas and Fan (1971) used an eddy cell model proposed by Lamont and Scott (1970) in conjunction with a wall model by Black (1969) and the time scale measured by Meek and Baer (1970, 1973) to model the whole process. In both these approaches, the differentiation between the instantaneous fluxes and their time-averaged values is unclear and rough approximations are necessary to effect closure of the solution. Experimental measurements to vindicate these visualisations are difficult to obtain because the wall layer in heat and particularly mass transfer processes is extremely thin. Perhaps the most extensive studies have been attempted by Hanratty and his associates. Their ideas have evolved, along with improved experimental evidence, from a belief that the eddy diffusivity near the wall is proportional to $y^4$ at very high Schmidt numbers (Son and Hanratty, 1967) as predicted by Deissler (1955), to a belief that a more accurate power index is 3.38 (Shaw and Hanratty, 1964, 1977) to an argument that the analogy between heat and mass transfer breaks down completely very close to the wall (Na and Hanratty, 2000). The research of Hanratty showed that the characteristic length scale of mass transfer in the longitudinal direction is equal to that for momentum transfer (Shaw and Hanratty 1964, 1977) but the time scale for mass transfer is much shorter than that for momentum transfer.

To explain this perplexing effect, Campbell and Hanratty (1983) have solved the unsteady mass transfer equations without neglecting the normal component of the convection velocity, which they model as a function of both time and distance. They found that only the low frequency components of the velocity fluctuations affect the mass transfer rates and that the energetic frequencies associated with the bursting process have no effect. In their explanation, the concentration sub-boundary layer acts as a low pass filter for the effect of velocity fluctuations on the mass transport close to the wall. The existence of two time scales in the wall region of heat or mass transfer has been noted by all modern investigators. Their explanation is varied. McLeod and Ponton (1977) differentiate between the renewal period and the transit time which is defined as the average time that an eddy takes to pass over a fixed observer at the wall. Loughlin et al. (1985) and more recently Fortuin et al. (1992) differentiate between the renewal time and the age of an eddy.

Trinh and Keey (1992a, 1992b) have shown that the boundary layer and penetration approaches to transport theory can be reconciled by the use of a time-space transformation that differentiates between two time scales: the contact time of the boundary layer flow and the diffusion time of heat across the boundary layer. Trinh (Trinh, 2009c) further discusses the significance of these time scales in turbulent transport.

## 1.3 Analogies between heat, mass and momentum transfer in non-Newtonian fluids

There are relatively few extensions of the analogies to purely viscous non-Newtonian fluids. Of the early attempts summarised by Skelland (1967) the most noticeable were those of Thomas (1960), Clapp (1961) and Metzner and Friend (1959). Further studies were made by Petersen and Christiansen (1966), McKillop (1972), Yoo (1974), Mishra and Tripathi (1973) Teng et al. (1979), Quader (1981), Quader and Wilkinson (1981) Kawase and Ulbrecht (1982, 1983), Wilson and Thomas (1985, 2006) Kawase and Moo-Young (1992). The correlation most often quoted in textbooks is that of Metzner and Friend who started with the derivation of Reichardt (1957, 1961)

$$St = \frac{(f/2)(\phi_v/\phi_h)}{1+(Pr-1)b\phi_v\sqrt{f/2}} \tag{15}$$

where

$$b = \int_0^{U^+} \frac{dU^+}{1+Pr(E_h/\nu)} \tag{16}$$

and $\phi_v, \phi_h$ are the ratios of maximum to average velocities and temperatures respectively.

Metzner and Friend used an approximate value of

$$\phi_v = 1.2 \tag{17}$$

and determined the function b experimentally to give

$$St = \frac{(f/2)}{1.2+(Pr-1)\sqrt{f/2}\left(Pr^{-1/3}\right)} \tag{18}$$

The results are often plotted in terms of the Nusselt number

$$Nu = \frac{hD}{k} \tag{19}$$

Which is related to the Stanton number by

$$Nu = St\left(\frac{D\rho C_p V}{k}\right) \tag{20}$$

For Newtonian fluids equation (20) may be rearranged as

$$Nu = St\,\text{Re}\,\text{Pr} \tag{21}$$

where $\text{Re} = \dfrac{DV\rho}{\mu}$ is the Reynolds number, and

$$\text{Pr} = \dfrac{C_p \mu}{k} \text{ the Prandtl number.}$$

Metzner and Friend first tried to apply equation (18) to non-Newtonian fluids by defining the Prandtl number as

$$\text{Pr}_g = \dfrac{C_p \mu_a}{k} \tag{22}$$

where

$$\mu_a = K \tau_w^{(n-1)/n} \tag{23}$$

is the apparent viscosity for a power law fluid but found that a better correlation of experimental data was obtained with a Prandtl number defined with what they call an "effective viscosity at the wall"

$$\text{Pr}_w = \dfrac{C_p \mu_w}{k} \tag{24}$$

$$\dfrac{\text{Pr}_g}{\text{Pr}_w} = \left(\dfrac{16}{\text{Re}_g\,f}\right)^{\frac{n'-1}{n'}} \left(\dfrac{3n'+1}{4n'}\right) \tag{25}$$

and

$$\text{Re}_g = \dfrac{D^{n'} V^{2-n'} \rho}{K 8^{n'-1} \left(\dfrac{3n'+1}{4n'}\right)^{n'}} \tag{26}$$

is the so called Metzner-Reed (1955) generalised Reynolds (MRRe) number. $D$ is the pipe diameter. Equation (18) correlated 80 data points with a standard deviation of 23.6% within the range

$$\dfrac{\text{Pr}_w \text{Re}_g}{n'^{.25}} \sqrt{\dfrac{f}{2}} > 5000 \tag{27}$$

We should note here that the apparent viscosity in equation (23) can only be calculated from experimental rheological data by assigning a rheological model to the fluid, in this case the power law

$$\tau = K \dot{\gamma}^n \tag{28}$$

The shear rate $\dot{\gamma}$ cannot be measured directly but must be derived from variables that can be measured. In laminar capillary flow the Mooney-Rabinowitsch equation allows the estimate of the wall shear rate as (Skelland, 1967)

$$\dot{\gamma}_w = \left(\frac{8V}{D}\right)\left(\frac{3n'+1}{4n'}\right) \qquad (29)$$

where $n' = \dfrac{d\ln(\tau_w)}{d\ln(8V/D)}$

In the general case a log-log plot of $\tau_w$ vs. $(8V/D)$ is a curve and $n'$ is not constant with varying shear stress or shear rate. The power law is only obeyed when this plot is a straight line in which case $n = n'$. Thus the use of $n'$ by the Metzner school to claim general application of their heat and momentum transfer data is a hybrid between a definition of apparent viscosity based on the power law and practical measurements of $n'$ at the shear stress measured in the transport experiments. Another doubt arises from the fact that equation (29) only applies to laminar flow. How can one then be sure that the value of $n'$ determined in a capillary viscometer in laminar flow would hold for the turbulent regime? As far as we know, the only proof that the flow curve $\tau_w$ vs. $\dot{\gamma}_w$ obtained with the Mooney-Rabinowitsch equation also applies to turbulent flow was presented in the unpublished work of Trinh (1969) and reported briefly in (Trinh, 2009c). With this proof we accept the use of $n'$ of Metzner and Friend which can at best be explained in terms of breaking up the $\tau_w$ vs. $(8V/D)$ curve into quasi linear sections.

Petersen and Christiansen (1966) claimed an improvement in the Metzner-Friend correlation with a modified version of the Prandtl number

$$\Pr_w = \frac{C_p \mu_w}{k}\left(\frac{\mathrm{Re}_{g,c}}{2100}\right) \qquad (30)$$

where $\mathrm{Re}_{g,c}$ is the critical value of $\mathrm{Re}_g$ at the end of the laminar regime, which is higher than the canonical critical Newtonian Reynolds number of $2100$.

## 2 Theory

## 2.1 The physical visualisation

The physical visualisation underpinning this paper and all others in this theory of turbulence is based on the wall layer process first illustrated dramatically through hydrogen bubble tracers by Kline et al. (1967) and confirmed by many others. In plan view, Kline et al. observed a typical pattern of alternate low– and high-speed streaks. The low-speed streaks tended to lift, oscillate and eventually eject away from the wall in a violent burst.

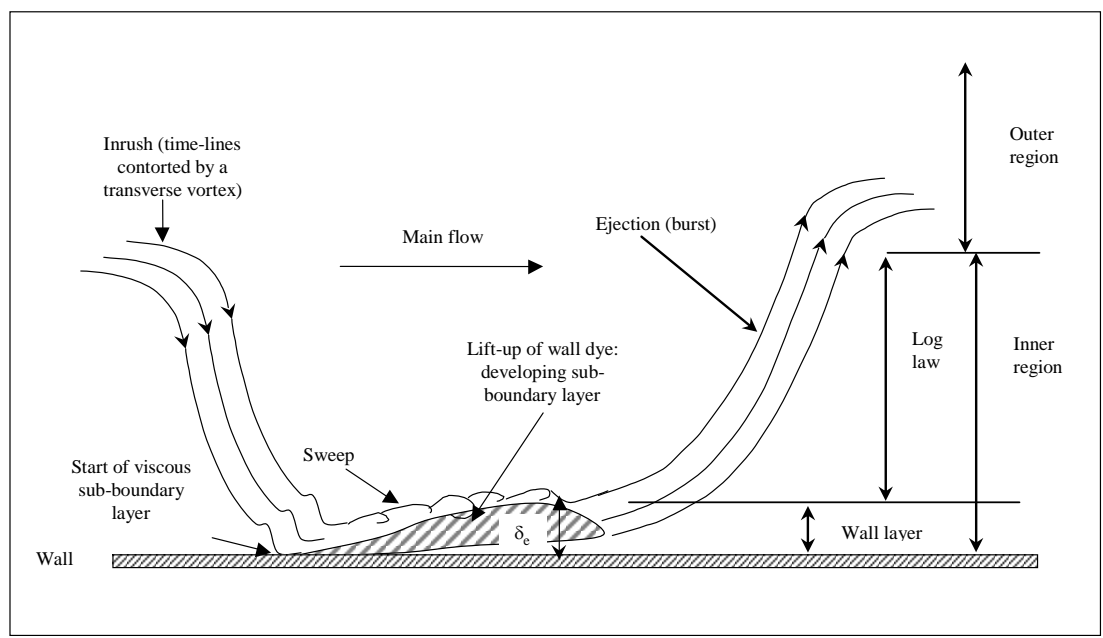

Figure 3 Schematic representation of the wall process in turbulent flow (Trinh, 2009c).

In side view, they recorded periodic inrushes of fast fluid from the outer region towards the wall. This fluid was then deflected into a vortical sweep along the wall. The low-speed streaks appeared to be made up of fluid underneath the travelling vortex. The bursts can be compared to jets of fluid that penetrate into the main flow, and get slowly deflected until they become eventually aligned with the direction of the main flow.

The sweep phase, which lasts longest and dominates the statistics of the flow near the wall, can be modelled with the method of successive approximations borrowed from the analysis of oscillating laminar boundary layers (Schlichting, 1960, Tetlionis, 1981). The first approximation, called the solution of order $\varepsilon^0$, describes the diffusion of viscous momentum into the main stream. The solution of order $\varepsilon$ and higher only become important when the fast periodic velocity fluctuations have become strong enough to induce jets of fluid to be ejected from the wall i.e. during the bursting phase (Trinh, 2009c).

In other words, the wall layer defined by the solution of order $\varepsilon^0$ is visualised as an unsteady state laminar sub-boundary layer which is interrupted by the emergence of the ejections. Mass, heat and momentum are contained in the same body of fluid ejected from the wall which explains, in our view, why there is an analogy between the laws of heat, mass and momentum in the outer region.

## 2.2  A theoretical derivation of the Metzner- Friend analogy

Trinh (1969) showed that the previous analogies could be summarised as the application of Reynolds' analogy from the outer region to a reference point within the wall layer. Three points are traditionally taken: the edge of the thermal wall layer $\delta_h^+, \theta_h^+$, the edge of the thermal buffer layer $\delta_{b,h}^+, \theta_{b,h}^+$ which was shown to be the time-averaged value of $\delta_h^+$ (Trinh, 2009c) and $\delta_{k,h}^+, \theta_{k,h}^+$ the edge of the diffusive sub-layers postulated by Prandtl. The form of the final correlation depends on the choice of the reference point: $\delta_{b,h}^+, \theta_{b,h}^+$ yields the Karman (1939) Martinelli (1947) analogies; $\delta_{k,h}^+, \theta_{k,h}^+$ the Prandtl-Taylor (1910) and Meztner-Friend (1958b) analogies.

Trinh and Keey (1992a, 1992b) showed that the solution of unsteady state diffusion of heat, mass and momentum can be formally transformed into steady state laminar boundary layer solutions and that the ratio of the thicknesses of the thermal and momentum boundary layers is equal to $Pr^{1/3}$ for $Pr > 5$ ..

The viscous laminar sub-layer was postulated by Prandtl as steady state laminar flow. The work of Kline et al. showed that such steady laminar flow does not exist in the wall layer of turbulent flows but the work of Trinh and Keey showed that unsteady state viscous and thermal diffusion do exist in turbulent flows and can be modeled by the relations (Trinh, 2009c):

$$\delta_{k,h}^+ = \delta_{k,v}^+ \Pr^{-1/3} \tag{31}$$

$$U^+ = y^+ \tag{32}$$

$$\theta^+ = y^+ \Pr \tag{33}$$

Hence

$$\theta_{k,h}^+ = U_{k,h}^+ \Pr \tag{34}$$

Equation (6) can be rearranged as

$$\theta_m^+ - \theta_{k,h}^+ = U_m^+ - U_{k,h}^+ \tag{35}$$

Now the Stanton number is related to the maximum normalised temperature by

$$\begin{aligned}\theta^+ &= \rho C_p \Delta \theta u_* / q_w \\ q_w &= h \Delta \theta_{av} \\ \theta_{av}^+ &= \frac{\theta_m^+}{\phi_h} = \frac{\rho C_p u_*}{h} = \left(\frac{\rho C_p V}{h}\right)\left(\frac{\tau_w}{\rho V^2}\right)^{1/2} = \frac{1}{St\phi_h}\sqrt{\frac{f}{2}}\end{aligned} \tag{36}$$

Rearranging equation (36) and taking account of equations (32) and (35)

$$\frac{1}{St\phi_h}\sqrt{\frac{f}{2}} = \frac{1}{\phi_v}\sqrt{\frac{2}{f}} + U_{k,v}^+(\Pr-1) \tag{37}$$

Thus

$$St = \frac{(f/2)(\phi_v/\phi_h)}{1+U_{k,h}^+\sqrt{f/2}(\Pr^{2/3}-1)} \tag{33}$$

In Newtonian fluids, the parameter $U_{k,v}^+$ is determined e.g. (Levich, 1962, Trinh, 1969) by the intersection of the well known log-law (Nikuradse, 1932)

$$U^+ = 2.5 \ln y^+ + 5.5 \tag{38}$$

and equation (33) giving

$$U_{k,v}^+ = 11.8 \tag{39}$$

Then

$$U_{k,h}^+ = 11.8 \Pr^{-1/3} \tag{40}$$

$$St = \frac{(f/2)(\phi_v/\phi_h)}{1 + \frac{11.8\sqrt{f/2}(Pr-1)}{Pr^{-1/3}}} \tag{41}$$

Trinh (1969) showed that for non-Newtonian fluids

$$U_{k,v}^{+} = 11.8\left(\frac{3n'+1}{4n'}\right) \tag{42}$$

Then

$$St = \frac{(f/2)}{1.2 + 11.8\left(\frac{3n'+1}{4n'}\right)\frac{(Pr_g - 1)}{Pr_g^{-1/3}}} \tag{43}$$

Equation (44) correlated 94 data points with $Re_g > 10000$ reported by Friend (1958) with a standard deviation of 24%. More accurate theoretical correlations are discussed in greater details in a previous work (Trinh, 1969, Trinh, 2009c).

## 2.3  A master curve based on the instantaneous wall shear stress

It is well known that the wall layer in non-Newtonian fluids appears thicker than in Newtonian fluids when normalised with the time averaged wall shear stress (Bogue and Metzner, 1963), consequently the friction factor plots fall on a family of lines (Dodge and Metzner, 1959) as shown in Figure 4. We have argued (Trinh, 2009b) that turbulence is an unsteady state process and must be correlated with local instantaneous parameters, not time averaged values. When the friction factor and Reynolds number were expressed in terms of the critical instantaneous shear stress at the point of ejection all the data collapsed into a single master curve (Figure 5). This indicated that the mechanism of turbulence production is the same in Newtonian and non-Newtonian fluids and the family of curves in Figure 4 was simply a consequence of an integration constant not accounted in previous solutions of the Reynolds equations.

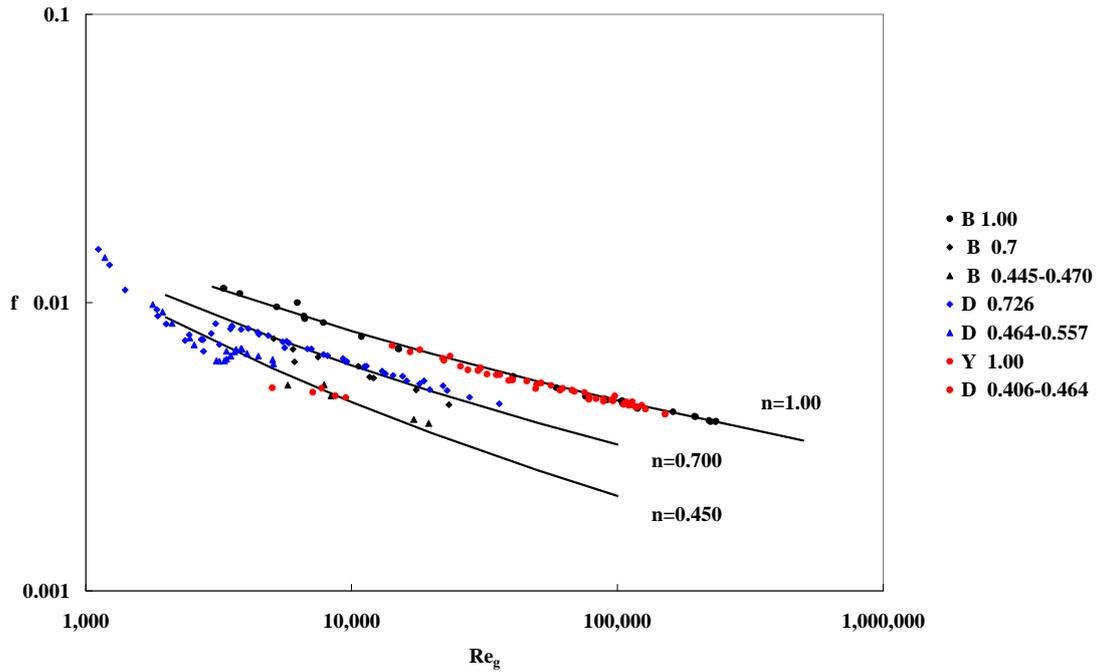

Figure 4 Friction factor for power law fluids using the MRRe number. From Trinh (2009b). Data from B (Bogue, 1962), D (Dodge, 1959), Y (Yoo, 1974).

The instantaneous critical shear stress at ejection is given by

$$\tau_w = (n+1)\tau_e \tag{44}$$

The apparent viscosity

$$\mu_e = K^{1/n}\tau_e^{(n-1)/n} \tag{45}$$

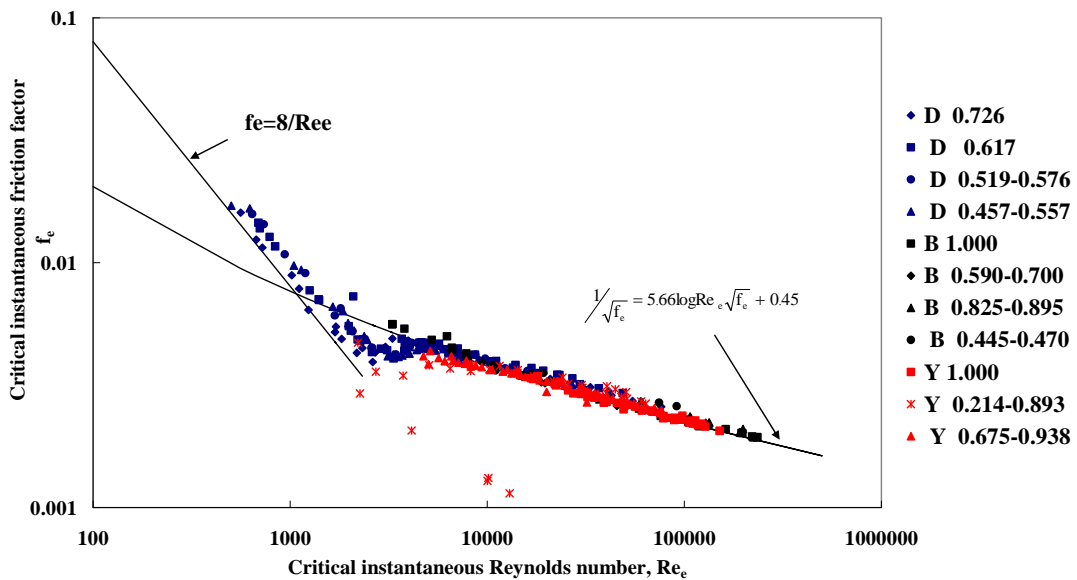

Figure 5 Plot of instantaneous friction factor and Reynolds number. From (Trinh, 2009b). Data source as in Figure 4.

Giving a Prandtl number

$$\Pr_e = \frac{\rho C_p \mu_e}{k} \qquad (46)$$

The widely used MRRe number

$$\operatorname{Re}_g = \frac{D^n V^{2-n} \rho}{K\, 8^{n-1} \left(\dfrac{3n+1}{4n}\right)^n} \qquad (47)$$

is related to the instantaneous Reynolds number by

$$\operatorname{Re}_e = \left(\operatorname{Re}_g f^{1-n}\right)^{\frac{1}{n}} 2^{\frac{5(n-1)}{n}} \left(\frac{3n+1}{4n}\right)\left(\frac{n+1}{2}\right)^{\frac{n-1}{n}} \qquad (48)$$

Since both momentum and heat transfer in turbulent flows are defined by the onset of ejections that convect wall fluids to the core region, we expect that a plot of Nusselt number against instantaneous Reynolds number will also collapse all data into a unique master curve.

## 3   Verification of theory

A plot of Nusselt number against MRRe number for different values of $n'$ is shown in Figure 6. There is clearly no particular pattern because not all differences in experimental conditions have been accounted for.

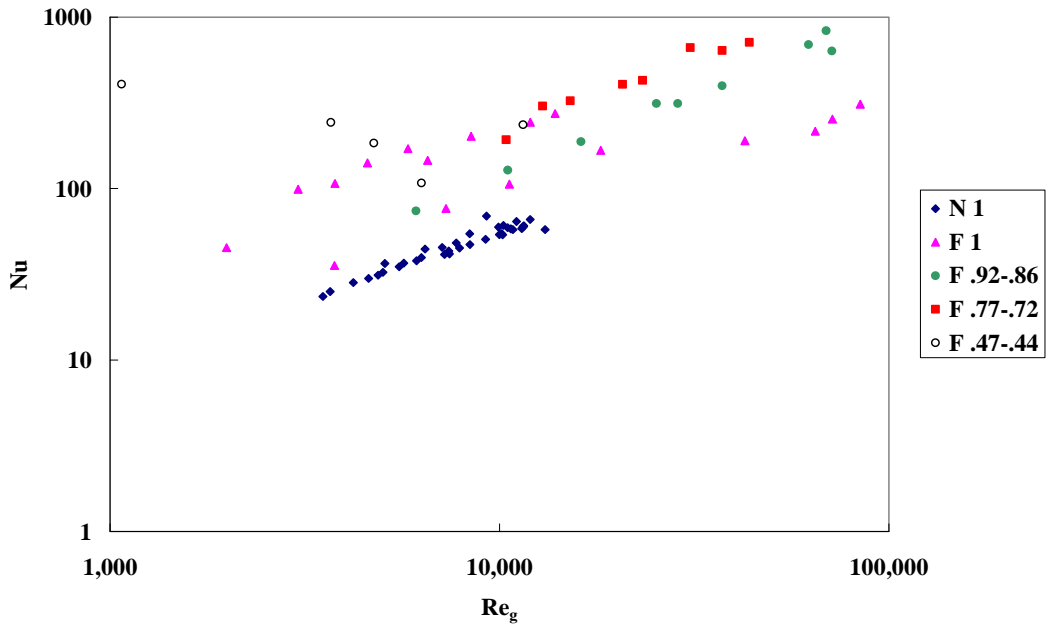

Figure 6 Plot of Nusselt number against MR Reynolds number. Data of F: Friend (1958) and N: Kiaka (2011).

A plot of $Nu/\Pr^{1/3}$ vs. $\mathrm{Re}_g$ in Figure 7 collapses the Newtonian data of Friend (1958) and Kiaka (2011) indicating the importance of accounting for the ratio of penetration thicknesses of heat and momentum diffusion in the wall layer.

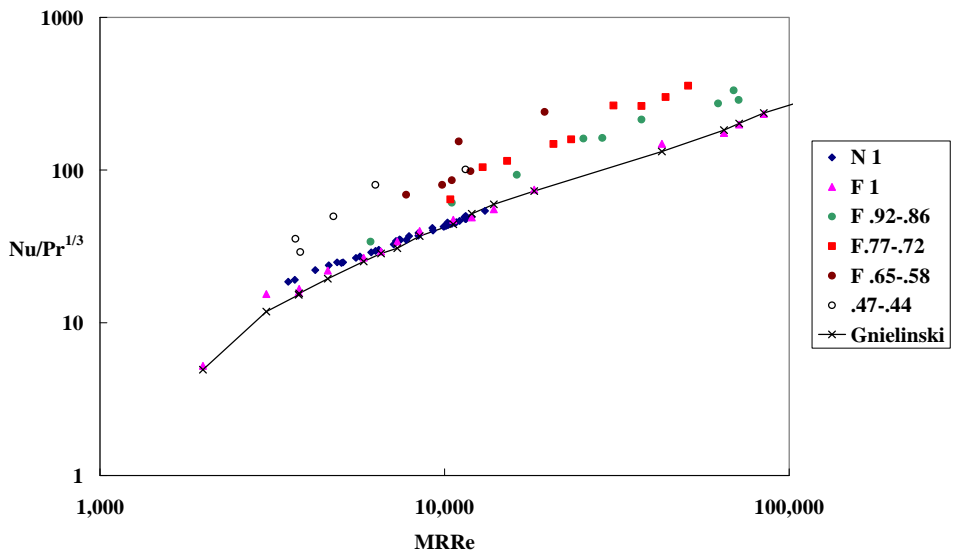

Figure 7 Plot of $Nu/\Pr^{1/3}$ vs. $\mathrm{Re}_g$. Data source as in Figure 6.

But the effect of the behaviour index $n'$ in power law fluids is still not accounted for. We have also indicated the goodness of fit of the Gnielinski correlation (1976) for the transition region in Newtonian fluids.

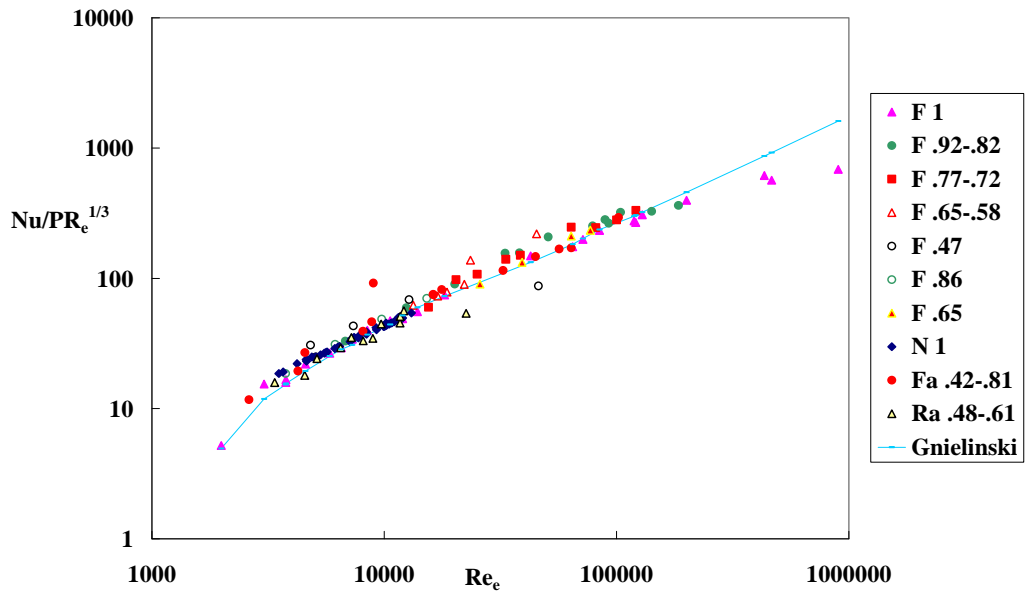

Figure 8  Plot of $Nu/\Pr_e^{1/3}$ vs. $\text{Re}_e$ for all $n'$. Data from F: Friend N: Kiaka, Fa: Farmer (1958), Ra: Raniere (1957).

When the instantaneous shear stress is used to evaluate the apparent viscosity, the data collapse clearly into a unique master curve as shown in Figure 8.

## 4  Discussion

The success of our data transformation should be viewed within the context of our integrated theory of turbulence (Trinh, 2009c). First of all, it emphasises again a key theme in our approach: that turbulence is an unsteady state phenomenon and needs to be modeled with the instantaneous parameters, not the time averaged values i.e. starting from the Navier-Stokes equations, not the Reynolds equations (Reynolds, 1895). The underlying philosophy of Reynolds is of course valid: practical engineers need to deal with time averaged values whereas mathematicians and more theoretical physicists and engineers are interested in turbulence mechanisms and the multitude of transient coherent structures that have been clearly identified e.g. (Cantwell, 1981, Robinson, 1991). Since the full Navier-Stokes equations are very difficult to solve, particularly with a four component decomposition of the local instantaneous velocity (Trinh, 2009a) a practical solution is to solve analytically a simplified subset of the Navier-Stokes equations and feeding the information into the Reynolds equations to effect closure (Trinh, 2010c). A second approach is to produce master curves that apply to all fluids and all geometries. This is achieved by recognising that the coherent structures, particularly the low speed streaks in the sweep phase and the ejections in the bursting phase define distinct areas in the turbulent flow fields, variously called regions, zones or layers that follow different rules. The normalising parameters for the master curves are either time-averaged values of velocity and distance at the interfaces of the zones (Trinh, 2010e, Trinh, 2010d) or the instantaneous ratios of kinetic and viscous energy at the interface of the sweep and bursting phases (Trinh, 2009b), this work. The advantage of this second approach (Trinh, 2009b) is that we do not need to give mathematical correlations for practical engineering problems. The user can simply use the master curve to read the relevant normalised parameters and back engineer the parameters required for situations of interest. Examples of this procedure have been given for friction factors.(Trinh, 2010a, Trinh, 2010b).

A major incentive for this work is its potential for food process operations since all three of us are engaged in teaching food technology and engineering. A large majority of food products are non-Newtonian, often with very high viscosities. Food process operations do not normally reach high turbulence, in fact some of our colleagues concentrate on teaching laminar correlations only. We believe that a substantial number of processes are operated in the transition region that is the most poorly understood of the three flow regimes. In fact it is well-known that the transition region in heat transfer is much larger than the transition region in friction factor plots, up to $Re = 10,000 - 20,000$ depending on the publication. Tam and Ghajar (2006) identified - correctly in our view - two major contributions by Churchill (1977) and Gnielinski (Gnielinski, 1976) for Newtonian fluids. The basic principles in these correlations is to interpolate between correlations for fully developed turbulent flow and values at the end of the laminar regime. Petersen and Christiansen (1966) have applied this approach to non-Newtonian pipe flow.

The correlation of Gnielinski, which has been quoted formally 375 times,

$$Nu = \frac{(f/2)(\text{Re}-1000)\text{Pr}}{1+12.7\sqrt{\frac{f}{2}}\left(\text{Pr}^{2/3}-1\right)} \qquad (49)$$

is clearly based on the correlation of Metzner and Friend. While even a cursory recasting of the Gnielinski correlation in terms of the instantaneous wall shear stress fits the data well, as shown in Figure 8, we prefer to leave formal correlations for the transition region until we have extended this master curve to include other fluid models, like the Bingham plastic and Herschel-Bulkley and other geometries. Clearly, much work remains to be done but we believe that we have made a good start.

## 5    Conclusion

A master curve for heat transfer to Newtonian and power law fluids has been successfully constructed based on the use of the critical instantaneous wall shear stress at the interface of the sweep and bursting phases of the wall process in turbulence production. The implications of the technique to understanding of turbulence fundamentals and practical applications have also been discussed.